\documentclass[doublespacing]{elsart}

\usepackage{graphicx}
\usepackage{amssymb}

\begin{document}
\begin{frontmatter}


\journal{SCES'2001: Version 1}


\title{Charge Stripe in an Antiferromagnet: 1d Band of Composite Excitations}

%
%
%
%
%
%

\author[ORNL]{A. L. Chernyshev\corauthref{1}},
\author[UCI]{Steven R. White},
\author[BU]{A. H. Castro Neto}

%
 
\address[ORNL]{Oak Ridge National Laboratory, 
P.O. Box 2008, Oak Ridge, TN 37831} 
\address[UCI]{Department of Physics, University of California, 
Irvine, CA 92697}
\address[BU]{Department of Physics, Boston University, Boston, MA 02215}

%
%
%
%

%
%
%
%

\corauth[1]{Corresponding Author: Solid State Division,
Oak Ridge National Laboratory,        
Bethel Valley Road, P.O. Box 2008,    
Bldg. 3025, MS-6032,                   
Oak Ridge, TN  37831-6032, Phone: 1-(865)-241-4325, Fax:
1-(865)-574-4143, Email: sasha@solid.ssd.ornl.gov}


\begin{abstract}

With the help of analytical and numerical studies of the $t$-$J_z$ model 
we argue that the charge stripe in an antiferromagnetic insulator
should be understood as a system of holon-spin-polaron excitations 
condensed at the self-induced antiphase domain wall. The structure of
such a charge excitation is studied in detail with numerical and
analytical results for various quantities being in a very
close agreement. An analytical picture of these excitations occupying 
an effective 1D stripe band is also in a very good accord with numerical
data. The emerging concept advocates the primary role
of the kinetic energy in favoring the stripe as a ground state. A 
comparative analysis suggests the effect of pairing and collective
meandering on the energetics of the stripe formation to be secondary. 

\end{abstract}

%
%



\end{frontmatter}

%
%
%
%
%

Unbiased numerical studies of strongly 
correlated models are a very important test of the ability of
theoretical approaches to describe the stripe phases in cuprates.
Importantly, pioneering studies  of the $t$-$J$ model 
on large clusters using the
Density Matrix Renormalization Group (DMRG) method
argue for the presence of stripes in the ground state \cite{WS}. 
However, some other numerical studies give
conflicting results on the very presence of stripe phases in this 
model, raising the question that the stripes seen in DMRG might be the
result of strong finite-size effects \cite{WS,HM}. 
Another aspect of the problem is that
numerical methods alone do not directly answer
questions on the origin and physics of stripes.
Ideally, one would wish for a theory which would closely agree 
with the numerical
data on all essential aspects, thus providing a
definite physical answer on how and why stripes are created.

In our work we present a study of the structure of a charge stripe 
embedded in an antiferromagnetic insulator 
within the $t$-$J_z$ model using analytical and
numerical approaches. Our numerical study utilizes DMRG 
in large $L_x\times L_y$ clusters up
to $11\times 8$ sites, using various boundary conditions. The analytical
study uses a self-consistent Green's function technique to account for
the retraceable-path motion of the holes away from the stripe. 
We demonstrate that the stripe in an antiferromagnet (AF)
should be viewed as a system of composite holon-spin-polaron excitations 
condensed at the self-induced antiphase domain wall (ADW). 

Our starting point is the $t$-$J_z$ model, which is given by:
\begin{eqnarray}
\label{H}
H = -t\sum_{\langle ij\rangle\sigma}(\tilde{c}^\dagger_{i\sigma}
\tilde{c}_{j\sigma}+{\rm H.c.})+ 
J\sum_{\langle ij \rangle} \bigl[S_i^z   S_j^z-
\frac{1}{4}N_iN_j \bigr]\ , 
\end{eqnarray}
where $t$ is the kinetic energy, $J$ is the AF
exchange, and $N_i=n_{i\uparrow}+n_{i\downarrow}$. 
All operators are defined in the space without
double-occupancy of the sites.
As the limiting case of the more general $t-J$
model, the $t-J_z$ model nevertheless captures the 
essential physics of holes in an AF. Technically, 
switching off the transverse spin fluctuation allows us to calculate
the effective 1D hole band in a highly controlled way. 
Thus, it is a
first-principles study of the stripe within a strongly correlated
model which goes far beyond the limits of mean-field or perturbation
theory.

The ``bare'' Green's function of the hole residing at the ADW
 is the Green's function of a free spinless fermion
(holon) with simple tight-binding dispersion.
The renormalization of this Green's function is important and, 
as in the case of the spin polaron,
is coming from the retraceable path movements of the hole away from
ADW and back. The retraceable path
approximation is equivalent to the self-consistent Born approximation
for the self-energy. 
The full Green's function is
then given by
\begin{equation}
\label{G}
G(k_y,\omega)=\frac{1}{\omega-2t\cos k_y-\Sigma(\omega)+i0}\
, 
\end{equation}
where $\Sigma(\omega)$ takes the form of a continued fraction,
\begin{equation}
\label{S}
\Sigma_{x_0}(\omega)=\frac{(z-2)t^2}{\omega-\omega_1-\frac{(z-1)t^2}
{\omega-\omega_1-\omega_2-\dots}}\ ,
\end{equation}
$\omega_i=J$ is the energy of the $i$-th segment of the string, which
is equal to the number of broken AF bonds ($J/2$ each) associated with
the segment. 
Since the energy spectrum of the elementary excitations is given
by the poles of the Green's function Eq. (\ref{G}), 
one needs to calculate $\Sigma(\omega)$ and seek 
solutions of $E(k_y)-2t\cos k_y-\Sigma(E(k_y))=0$. 
The resulting effective 1D band for the composite holon-spin-polaron
excitation has been calculated in our previous work \cite{CCNB}.

Here we present the results of the DMRG calculations for the cluster
$11\times8$ with mixed boundary conditions. Namely, the open boundary
conditions are applied in the wide direction of the cluster with the
staggered field in order to enforce the ADW inside the system. The
periodic boundary condition of the special type (M\"obius boundary
conditions, Ref. \cite{us}) are
applied in the narrow direction of the cluster in order 
to avoid the frustration of the holonic motion along
the stripe. The results for the ground state energy as a function of
$t/J$ for such a system are shown 
in Fig. 1
together with the results of analytical approach for 
the single composite excitation described above. 
The agreement between numerical and analytical results is close to the
accuracy of the numerical data and is even better than for the case
without an ADW. This is because the so-called Trugman corrections to
the retraceable path approximation in our case are suppressed due to a
conservation of holonic quantum numbers in the virtual decay processes.

Another important comparison is made for the systems with variable
number of holes, that is at the different occupation fraction of the
effective 1D band. Assuming a ``rigid band'' filling of the stripe
band within our theoretical approach we calculate the total energy of
the system per hole. This result can be compared to the same quantity
obtained by DMRG in $11\times8$ cluster. The total energy per hole
versus the 1D hole density from numerical and analytical approaches
for $J/t=0.35$ are shown in Fig. 2.
The rigid band approximation neglects the effects of the
stripe meandering (collective motion of the excitations) and the
effects of holon-holon (polaron-polaron) interactions.
However, given the good 
agreement of analytical and numerical results
 one can conclude
that such effect are secondary for the stripe formation. Therefore,
the kinetic energy of the holes, both along the stripe (holonic
motion) and perpendicular to it (spin-polaron part), is the main 
reason which brings the stripe to the ground state.

Summarizing, we have presented a comparison of the DMRG numerical data
for a large cluster with the self-consistent Green's function
based analytical studies of the single stripe of holes in an
AF described by the
$t$-$J_z$ model. Close agreement of the results gives strong 
support to the validity of our analytical method.
Longitudinal as well as transverse
kinetic energy of holes are explicitly taken into account in our
approach, and their role in stabilizating the stripe
as a ground state of the system is revealed. We have provided a description
of the charge carriers building the stripe as a system of 1D
elementary excitations, unifying the features of holons, spinons, and AF
spin polarons. This represents a new level of 
 understanding of the structure of the stripe phase in cuprates.
Further work on this subject will be reported at a later time.

This research was supported in part by 
Oak Ridge National Laboratory, 
managed by UT-Battelle, LLC, for the U.S. Department of Energy under
contract DE-AC05-00OR22725, and by a
CULAR research grant under the auspices of the US Deparment of Energy.

%
%
%
%

%
%
 \begin{figure}
     \centering
     \includegraphics{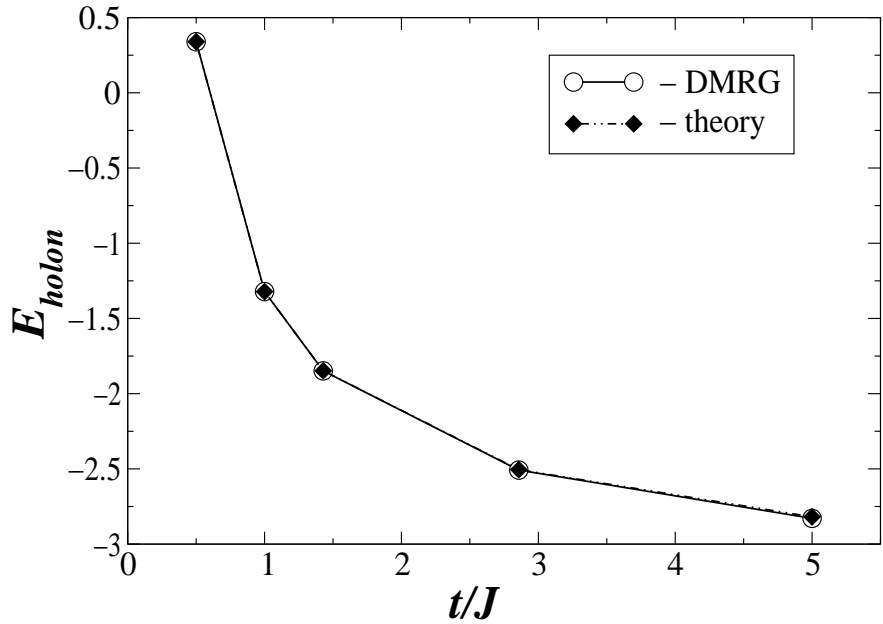}
     \caption{The ground state energy relative to the energy of the
     empty system in units of $t$ as a function of
     $t/J$ for $11\times 8$ cluster (circles) and analytical results
     for the energy of the single holonic-spin-polaron (diamonds).}
 \end{figure}  
 \begin{figure}
     \includegraphics{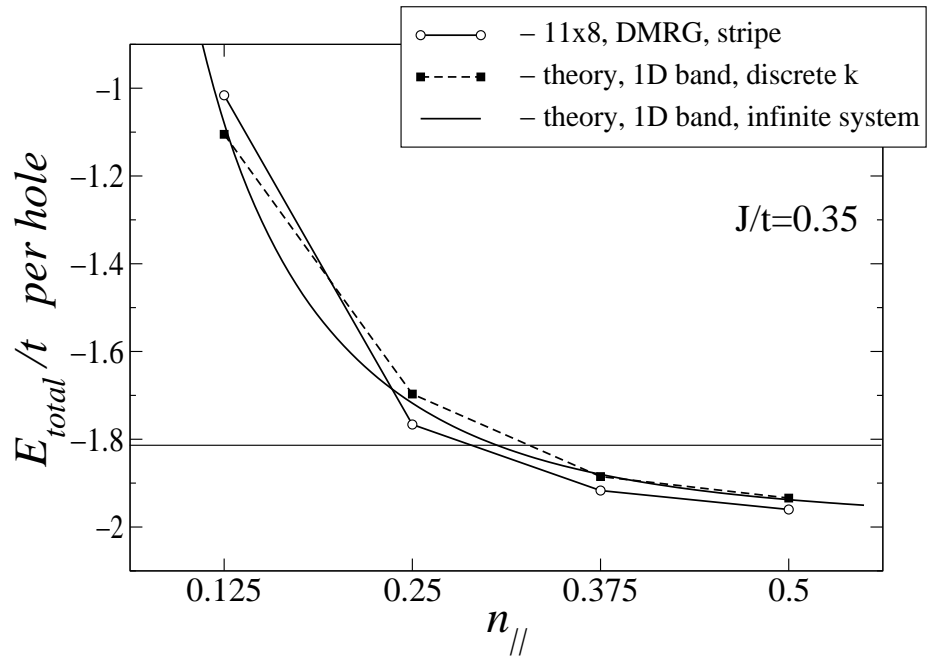}
     \caption{The total energy of the system relative to the energy of the
     empty system per hole
versus the 1D hole density from numerical (circles) and analytical
(squares, solid line) approaches.}
 \end{figure}  
%
%


\end{document}